\documentclass[3p,times]{elsarticle}

\usepackage{ecrc}

\volume{00}

\firstpage{1}

\journalname{Nuclear Physics A}

\runauth{}

\jid{nupha}

\usepackage{amssymb}
\usepackage{url}

\usepackage[figuresright]{rotating}

\begin{document}

\begin{frontmatter}

%% Title, authors and addresses

%% use the tnoteref command within \title for footnotes;
%% use the tnotetext command for the associated footnote;
%% use the fnref command within \author or \address for footnotes;
%% use the fntext command for the associated footnote;
%% use the corref command within \author for corresponding author footnotes;
%% use the cortext command for the associated footnote;
%% use the ead command for the email address,
%% and the form \ead[url] for the home page:
%%
%% \title{Title\tnoteref{label1}}
%% \tnotetext[label1]{}
%% \author{Name\corref{cor1}\fnref{label2}}
%% \ead{email address}
%% \ead[url]{home page}
%% \fntext[label2]{}
%% \cortext[cor1]{}
%% \address{Address\fnref{label3}}
%% \fntext[label3]{}

\dochead{}
%% Use \dochead if there is an article header, e.g. \dochead{Short communication}

\title{Modeling the Impact Parameter Dependence of the nPDFs With EKS98 and EPS09 Global Fits}

%% use optional labels to link authors explicitly to addresses:
%% \author[label1,label2]{<author name>}
%% \address[label1]{<address>}
%% \address[label2]{<address>}

%\author[jyu,hip]{I.~Helenius\corref{cor1}}
\author[jyu,hip]{I.~Helenius}
%\author[jyu,hip]{I.~Helenius (speaker)}
\ead{ilkka.helenius@jyu.fi}
\author[jyu,hip]{K.J.~Eskola}
%\ead{kari.eskola@phys.jyu.fi}
\author[psu]{H.~Honkanen}
%\ead{hmh17@psu.edu}
\author[sdg,cern]{C.A.~Salgado}
%\ead{carlos.salgado@usc.es}

%\cortext[cor1]{Speaker}

\address[jyu]{Department of Physics, P.O. Box 35, FI-40014 University of Jyv\"askyl\"a, Finland}
\address[hip]{Helsinki Institute of Physics, P.O. Box 64, FIN-00014 University of Helsinki, Finland}
\address[psu]{The Pennsylvania State University, 104 Davey Lab, University Park, PA 16802, USA}
\address[sdg]{Departamento de F\'\i sica de Part\'\i culas and IGFAE, Universidade de Santiago de Compostela, Galicia-Spain}
\address[cern]{Physics Department, Theory Unit, CERN, CH-1211 Gen\`eve 23, Switzerland}

\begin{abstract}
%% Text of abstract
So far the nuclear PDFs (nPDFs) in the global DGLAP fits have been taken to be spatially independent. In this work \cite{Helenius:2012wd}, using the $A$-dependence of the globally fitted sets EPS09 and EKS98, we have determined the spatial dependence of the nPDFs in terms of powers of the nuclear thickness functions. New spatially dependent nPDF sets EPS09s (NLO, LO, error sets) and EKS98s (LO) are released. As an  application, we consider the nuclear modification factor $R_{\rm dAu}^{\pi^0}$ at midrapidity for neutral pion production in deuteron-gold collisions at RHIC in NLO. Comparison with the PHENIX data in different centrality classes is also shown. In addition, predictions for the corresponding nuclear modification factor $R_{\rm pPb}^{\pi^0}$ in proton-lead collisions at the LHC are discussed. 
\end{abstract}

\begin{keyword}
Nuclear PDFs \sep hard processes \sep centrality dependence \sep d+$A$ collisions \sep p+$A$ collisions
%% keywords here, in the form: keyword \sep keyword

%% MSC codes here, in the form: \MSC code \sep code
%% or \MSC[2008] code \sep code (2000 is the default)

\end{keyword}

\end{frontmatter}

%%
%% Start line numbering here if you want
%%
% \linenumbers

%% main text
%\section{ }
%\label{}

\section{Introduction}

In a high-energy hadronic or nuclear collision of particles $A$ and $B$ the inclusive cross sections for hard processes, where the interaction scale is large, $Q^2\gg\Lambda^2_{\rm QCD}$, can be computed using the QCD collinear factorization theorem \cite{Collins:1989gx,Brock:1993sz}, 
\begin{equation}
\mathrm{d} \sigma^{AB \rightarrow k + X} = \sum\limits_{i,j,X'} f_{i}^A(Q^2) \otimes f_{j}^B(Q^2) \otimes  \mathrm{d}\hat{\sigma}^{ij\rightarrow k + X'} + {\cal O}(1/Q^2), 
\label{eq:sigmaAB1}
\end{equation}
where $\mathrm{d}\hat{\sigma}$ are the perturbatively computable partonic pieces (cross sections in lowest order), and $f_i^A$ ($f_j^B$) is the parton distribution function (PDF) for a parton flavor $i$ ($j$) in $A$ ($B$).%(and correspondingly for the flavor $j$ in $B$).

As is well known, the PDFs of nucleons bound to a nucleus, the nPDFs, are modified relative to the free-nucleon PDFs. Global DGLAP analyses, analogous to the free-proton case, have been developed also for the nPDFs. In this study we have considered two globally fitted sets, the EKS98 \cite{Eskola:1998df} and EPS09 \cite{Eskola:2009uj}, where the nPDFs are defined in terms of the nuclear modification $R_{i}^{A}(x,Q^2)$ and the free nucleon PDF $f_i^N(x,Q^2)$ as
\begin{equation}
f_{i}^{A}(x,Q^2) = R_{i}^{A}(x,Q^2) \, f_{i}^{N}(x,Q^2)
\end{equation}
for each parton flavor $i$. Using the error sets of the EPS09 analysis one can also quantify the propagation of the nPDF uncertainties into the hard-process cross sections.
%the remaining uncertainties in the nPDFs. 
So far all these globally analyzed nPDFs have been spatially independent. Thus, it has not been possible to compute the nuclear hard-process cross-sections in different centrality classes consistently with the global analyses. This problem is adressed in \cite{Helenius:2012wd}, where the spatial dependence of the EPS09 and EKS98 nPDFs has been determined using the framework discussed in the next section.

\section{Analysis Framework}

\begin{figure}[bht]
\begin{minipage}[t]{0.49\linewidth}
\centering
\includegraphics[trim = 0pt 4pt 0pt 2pt, clip, width=\textwidth]{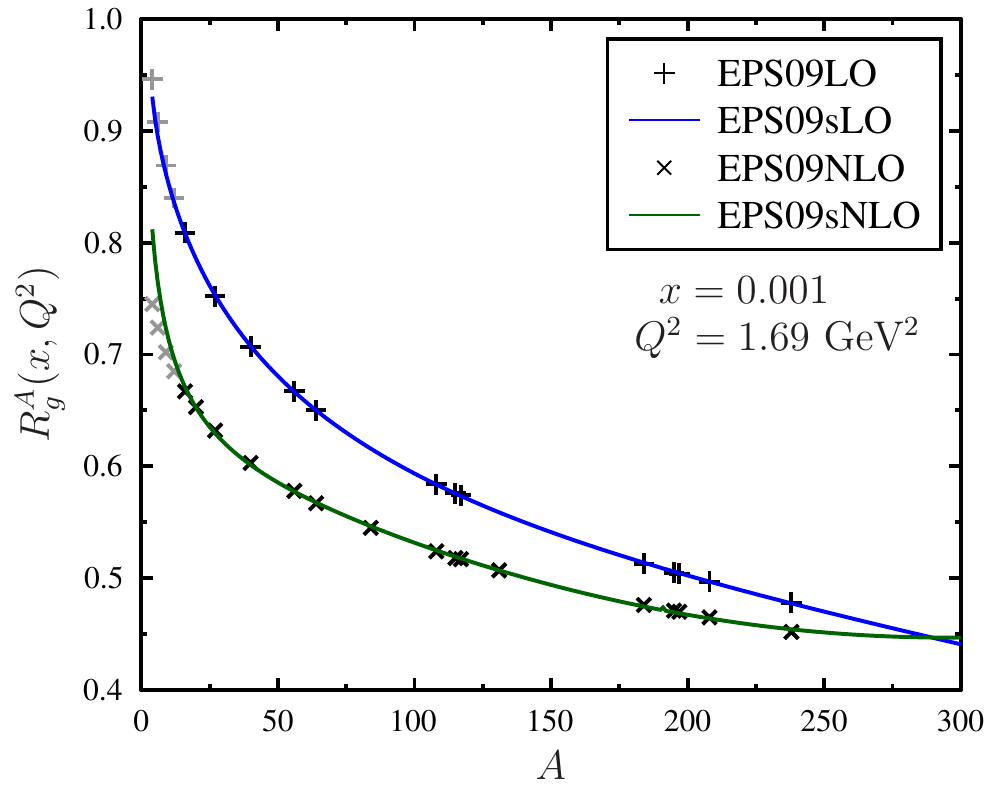}
\caption{The spatially averaged gluon modification $R_g^A(x,Q^2)$ at fixed values  $x=0.001$ and $Q^2=1.69$~GeV$^2$ as a function of $A$ from the central sets EPS09NLO (crosses) and EPS09LO (pluses) and from the corresponding spatial fits EPS09sNLO (green) and EPS09sLO (blue). The nuclei  at $A < 16$ (gray markers) were not used in our fits. From \cite{Helenius:2012wd}.}
\label{fig:R_g_A_fits}
\end{minipage}
\hspace{0.02\linewidth}
\begin{minipage}[t]{0.49\linewidth}
\centering
\includegraphics[trim = 20pt 18pt 0pt 15pt, clip, width=\textwidth]{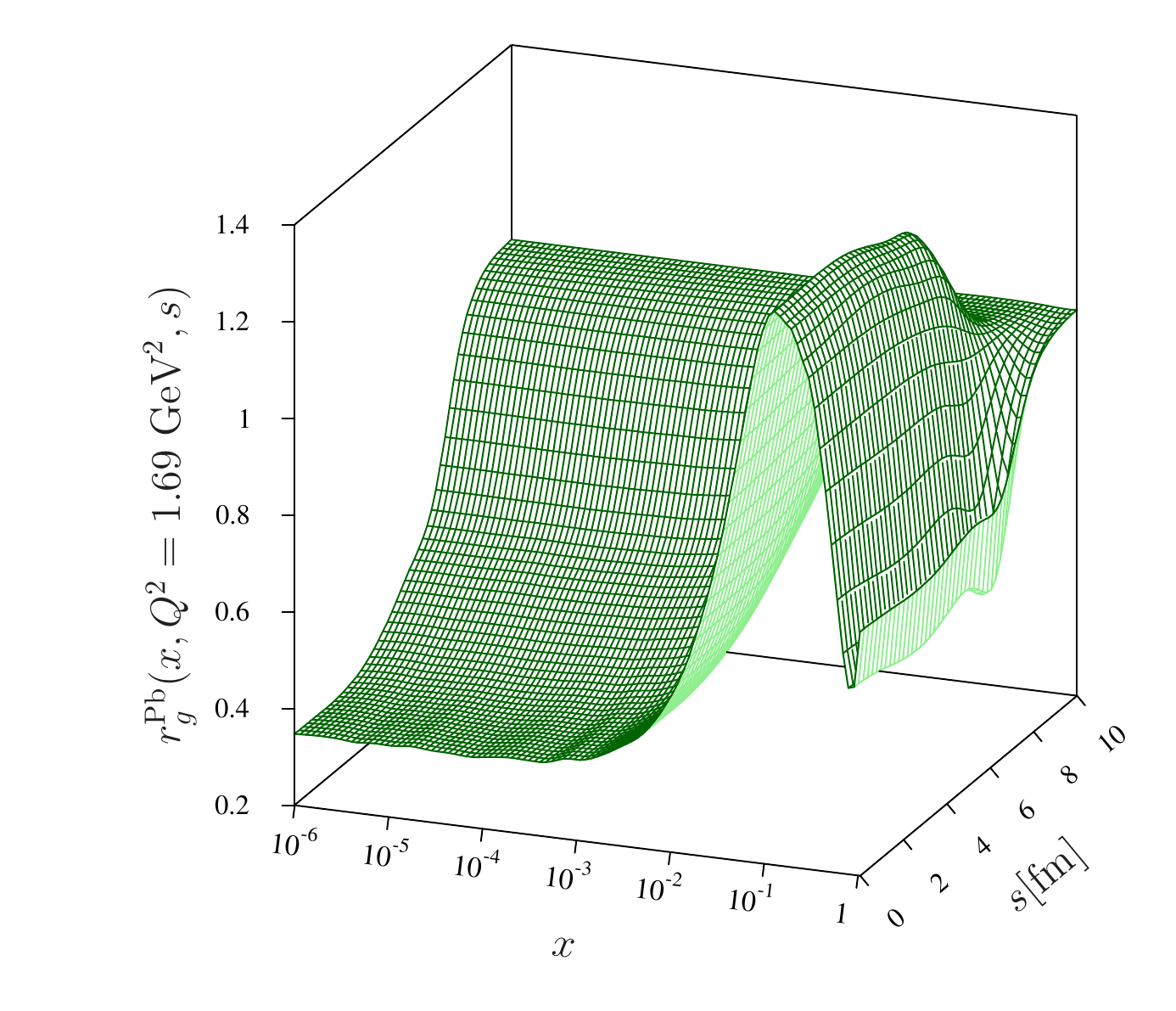}
\caption{The spatially dependent modification of gluon distribution in a Pb nucleus, $r_g^{\rm Pb}(x,Q^2,s)$, from EPS09sNLO as a function of $x$ and $s=|\mathbf{s}|$ at the initial scale $Q^2 = 1.69$~GeV$^2$. From \cite{Helenius:2012wd}.}
\label{fig:R_g_3d}
\end{minipage}
\end{figure}

%We introduce a nuclear modification $r_i^A(x,Q^2,\mathbf{s})$ which, in addition to $x$ and $Q^2$, depends also on the transverse position $\mathbf{s}$ of the nucleon inside the nucleus and define it so that spatial average of it yields the spatially independent nuclear modification:
We introduce a nuclear modification $r_i^A(x,Q^2,\mathbf{s})$ which, in addition to $x$ and $Q^2$, depends also on the transverse position $\mathbf{s}$ of the nucleon inside the nucleus. We define this quantity so that its spacial average yields the original (impact parameter independent) nuclear modification:
%\vspace{-10pt}
\begin{equation}
R_{i}^{A}(x,Q^2) \equiv \frac{1}{A}\int \mathrm{d}^2 \mathbf{s} \,T_A(\mathbf{s}) r_{i}^{A}(x,Q^2,\mathbf{s}),
\label{eq:series}
\end{equation}
where $R_{i}^{A}(x,Q^2)$ is obtained from the global analysis EKS98 or EPS09. 
The key assumption is that the spatial dependence of $r_i^A(x,Q^2,\mathbf{s})$ is a function of the nuclear thickness $T_A(\mathbf{s})$. The functional form we adopt and test here is a simple power series of the thickness function,
%\vspace{-10pt}
\begin{equation}
r_A(x,Q^2,\mathbf{s}) = 1 + \sum_{j=1}^{n} c_j^i(x,Q^2)\left[ T_A(\mathbf{s})\right]^j.
\end{equation}
%The fit parameters $c_i$ are now independent of $A$. We exploit the $A$-dependence of the spatially independent nPDFs to determine the values for these fit parameters. 
The $A$-independent parameters $c_j^i(x,Q^2)$ are obtained by fitting the $R_i^A(x,Q^2)$ in Eq.~\ref{eq:series} to the $A$-dependence of the spatially independent nPDFs.
In practice, we minimize the $\chi^2$ defined as
\begin{equation}
\chi^2_i(x,Q^2) = \sum_A \left[\frac{R^{A}_{i}(x,Q^2) - \frac{1}{A}\int \mathrm{d}^2 \mathbf{s}\, T_A(\mathbf{s})r^{A}_{i}(x,Q^2,\mathbf{s})}{W^{A}_{i}(x,Q^2)} \right]^2,
\end{equation}
where the weights $W^{A}_{i}(x,Q^2)$ are set by hand (see \cite{Helenius:2012wd}).
As can be seen from Fig.~\ref{fig:R_g_A_fits}, we can reproduce the $A$-dependence very well with the power series ansatz when we take into account terms up to the fourth order. This holds also for other parton flavors in the whole kinematic region considered and also for the error sets in EPS09 LO and NLO analyses. As an outcome of this fitting procedure we obtain the spatially dependent nPDF sets, which we refer to as \texttt{EPS09s} and \texttt{EKS98s} (''\texttt{s}'' for ''spatial'').
These sets are now available at our webpages\footnote{\url{https://www.jyu.fi/fysiikka/en/research/highenergy/urhic/nPDFs}}. As an example, the gluon modification for the lead nucleus ($A=208$) is shown in Fig.~\ref{fig:R_g_3d} as a function of $x$ and $s$. We notice that the nuclear modifications at small $s$ are larger than the spatially averaged nuclear modifications, and at larger $s$ values ($s \gg R_A$), the nuclear effects die out.

\section{Applications}

\begin{figure}[bth!]
\begin{center}
\includegraphics[trim = 0pt 7pt 0pt 15pt, clip, width=0.9\textwidth]{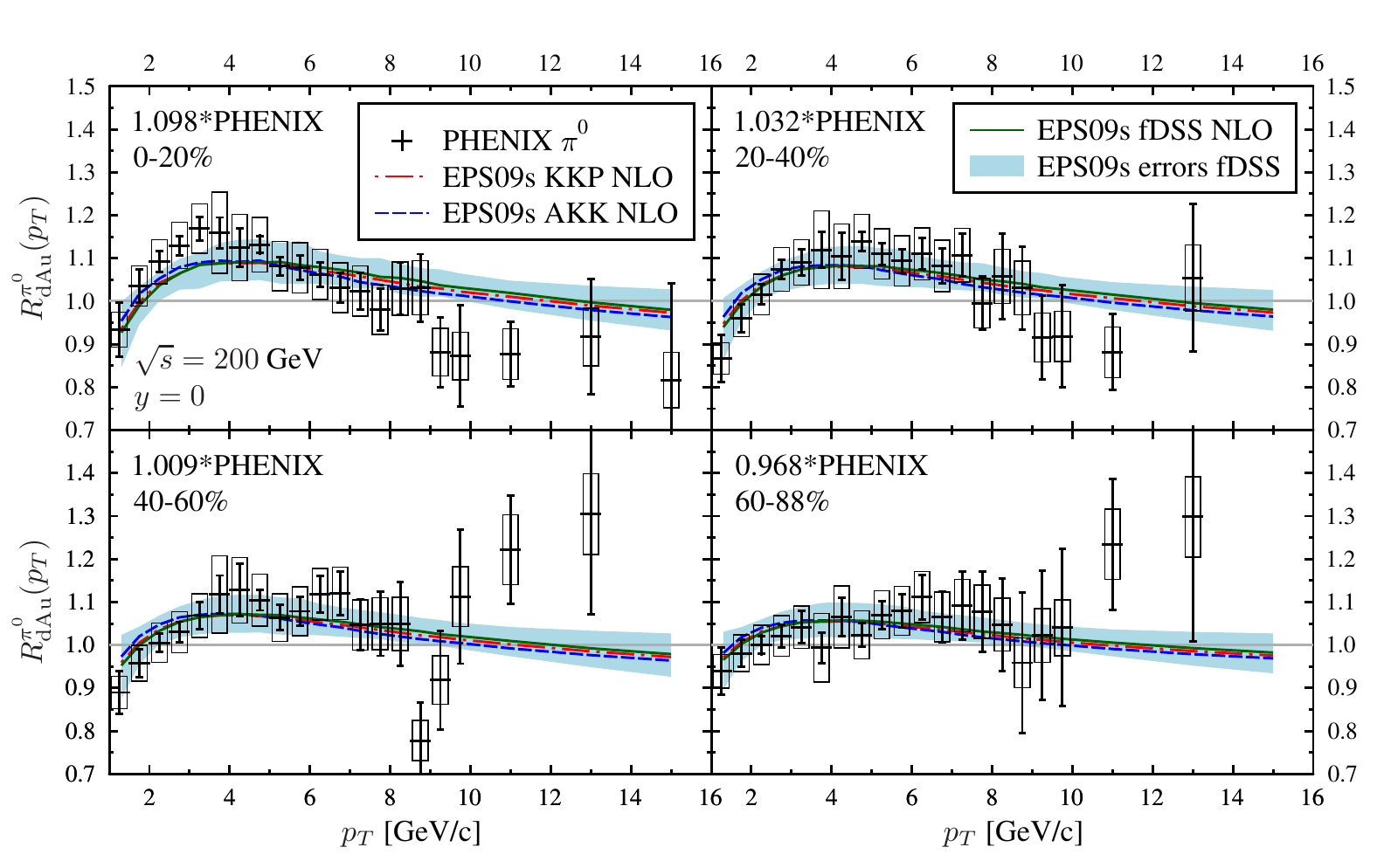} 
\caption{The nuclear modification factor $R^{\pi^0}_{\rm dAu}(p_T)$ in d+Au collisions for $\sqrt{s_{NN}} = 200$~GeV at $y=0$ for different centrality classes. Calculations are in NLO pQCD using EPS09s and three different fragmentation functions. The blue error bands are computed with the EPS09s error sets and fDSS, and the data are from PHENIX~\cite{Adler:2006wg}. From \cite{Helenius:2012wd}.}
\label{fig:R_dAuNLO}
\end{center}
\vspace{-10pt}
\end{figure}

As we are now equipped with these spatially dependent nPDFs, we are able to calculate the cross sections of hard processes also in different centrality bins. The nuclear modification factor for the production of a particle $k$ in a hard process for a given centrality class can be calculated from (see \cite{Helenius:2012wd} for details)
\begin{equation}
R_{AB}^{k}(p_T,y; b_1,b_2) %\equiv \frac{\left\langle\frac{\mathrm{d}^2 N_{AB}^{k}}{\mathrm{d}p_T \mathrm{d}y}\right\rangle_{b_1,b_2}}{\langle N_{bin}^{AB} \rangle_{b_1,b_2} \frac{1} {\sigma^{NN}_{inel}}\frac{\mathrm{d}^2\sigma_{\rm pp}^{k}}{\mathrm{d}p_T \mathrm{d}y}} 
= \frac{\int_{b_1}^{b_2} \mathrm{d}^2 \mathbf{b} \frac{\mathrm{d}^2 N_{AB}^{k}(\mathbf{b})}{\mathrm{d}p_T \mathrm{d}y} }{ \int_{b_1}^{b_2} \mathrm{d}^2 \mathbf{b} \,T_{AB}(\mathbf{b})\frac{\mathrm{d}^2\sigma_{\rm pp}^{k}}{\mathrm{d}p_T \mathrm{d}y}},
\label{eq:R_AB}
\end{equation}
where $T_{AB}(\mathbf{b})$ is the standard nuclear overlap function, and the impact parameter values $b_1$ and $b_2$ 
%and the average number of binary collisions 
for each centrality class can be calculated using the optical Glauber model \cite{Miller:2007ri}. In Fig.~\ref{fig:R_dAuNLO} we plot the nuclear modification factor $R_{\rm dAu}^{\pi^0}$ for neutral pion production in d+Au collisions at RHIC at midrapidity in four different centrality bins, 0-20~\%, 20-40~\%, 40-60~\% and 60-88~\%. For the thickness function of deuterium, see Ref.~\cite{Helenius:2012wd}.
The different lines correspond to NLO calculations (using \verb=INCNLO=-package\footnote{\url{http://lapth.in2p3.fr/PHOX_FAMILY/readme_inc.html}} \cite{Aversa:1988vb})
with three different fragmentation functions and the uncertainty band is computed using the error sets in EPS09s. For the free proton PDFs we use the CTEQ6M set \cite{Pumplin:2002vw} and all the scales have been fixed to $p_T$. The calculations are compared with the PHENIX data~\cite{Adler:2006wg}, which we have multiplied by a different overall factor in each centrality bin. These factors are well consistent with the overall normalization uncertainties quoted by the experiment. In all centrality classes the calculations seem to agree with the data, when all the uncertainties in the data and in the calculation are taken into account. Especially the evolution of the slope at the data in the small-$p_T$ region is well reproduced from central to peripheral collisions.

\begin{figure}[htb!]
\begin{center}
\includegraphics[trim = 0pt 7pt 0pt 15pt, clip, width=0.9\textwidth]{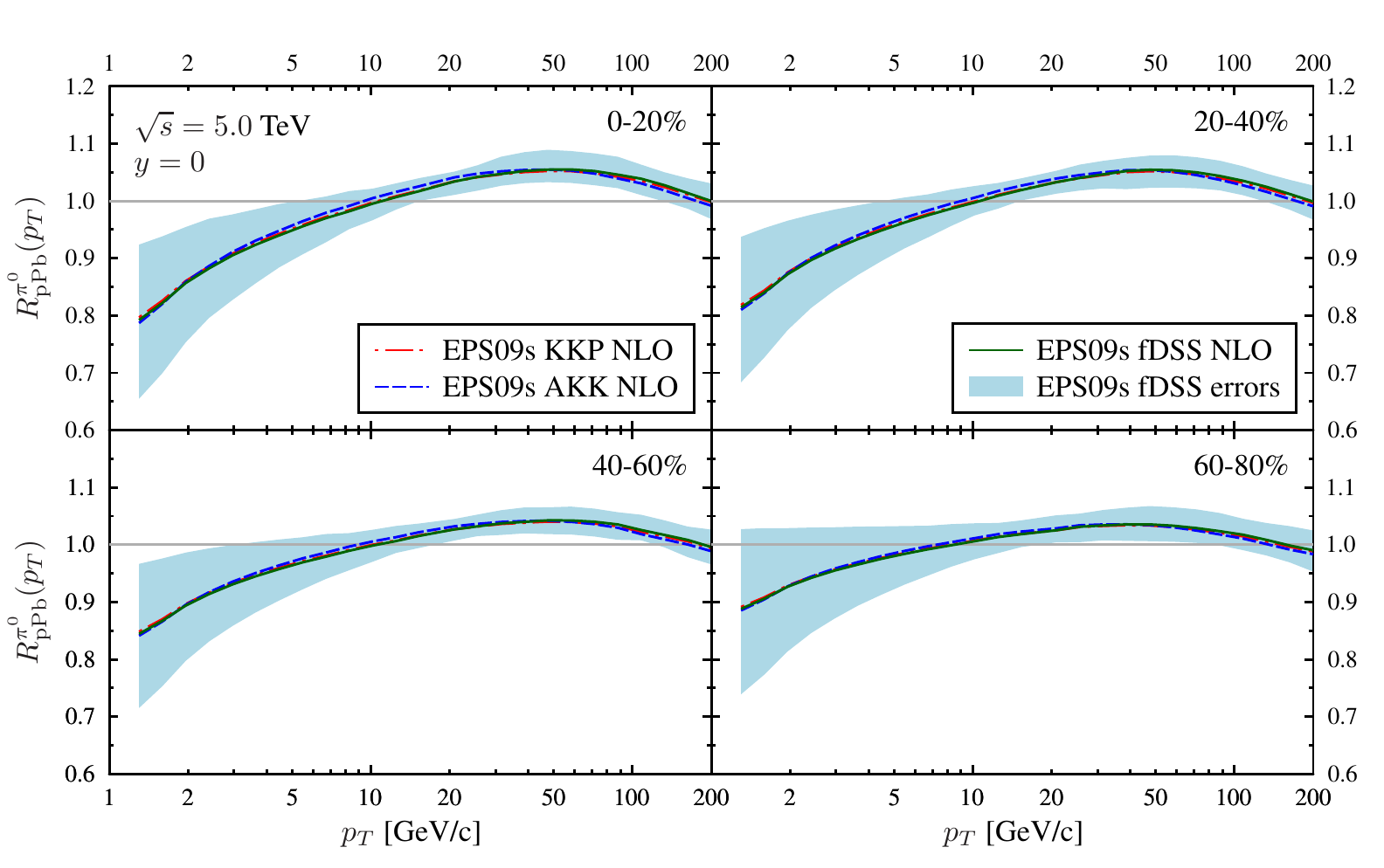}
\caption{The nuclear modification factor $R^{\pi^0}_{\rm pPb}(p_T)$ in p+Pb collisions for $\sqrt{s_{NN}} = 5.0$~TeV at $y=0$ for different centrality classes. Calculations are in NLO pQCD using EPS09s and three different fragmentation functions. The blue error bands are computed with the EPS09s error sets and fDSS \cite{deFlorian:2007aj}. From \cite{Helenius:2012wd}.}
\label{fig:R_pPb_pi0_y0_NLO}
\end{center}
\end{figure}

We have also performed the corresponding calculations for the forthcoming p+Pb collisions at the LHC and these results are shown in Fig.~\ref{fig:R_pPb_pi0_y0_NLO}. Due to the larger $\sqrt{s_{NN}}$, we get more shadowing in the small $p_T$ region than for d+Au collisions at RHIC. Also the basic features of our spatially dependent nPDFs are nicely visible: the nuclear effects are strongest in central collisions and they decrease towards peripheral collisions, and the difference between the central and the minimum bias collisions is rather small (see Ref.~\cite{Helenius:2012wd}).

\vspace{-10pt}

\section*{Acknowledgements}
{\small 
I.H. and K.J.E. thank the Magnus Ehrnrooth Foundation, Academy of Finland (Project 133005) and the PANU graduate school for financial support.
C.A.S. is supported by the European Research Council grant HotLHC ERC- 2001-StG-279579 and by Ministerio de Ciencia e Innovaci\'on of Spain, and is a Ram\'on y Cajal researcher.
H.H. is supported by the U.S. Department of Energy under Grant  DE- FG02-93ER40771.}

%\vspace{-5pt}

%% The Appendices part is started with the command \appendix;
%% appendix sections are then done as normal sections
%% \appendix

%% \section{}
%% \label{}

%% References
%%
%% Following citation commands can be used in the body text:
%% Usage of \cite is as follows:
%%   \cite{key}         ==>>  [#]
%%   \cite[chap. 2]{key} ==>> [#, chap. 2]
%%

%% References with BibTeX database:

%\bibliographystyle{elsarticle-num}
\bibliographystyle{model1a-num-names}
\bibliography{HP2012_proceedings}

%% Authors are advised to use a BibTeX database file for their reference list.
%% The provided style file elsarticle-num.bst formats references in the required Procedia style

%% For references without a BibTeX database:

% \begin{thebibliography}{00}

%% \bibitem must have the following form:
%%   \bibitem{key}...
%%

% \bibitem{}

% \end{thebibliography}

\end{document}